\documentclass[preprint,longabstract]{aastex}

\usepackage{amsmath}

\shorttitle{Circulation regimes of synchronous planets}
\shortauthors{Haqq-Misra et al.}

\begin{document}


\title{Demarcating circulation regimes of synchronously rotating terrestrial planets within the habitable zone}


\author{Jacob Haqq-Misra\altaffilmark{1,2}, 
Eric. T. Wolf\altaffilmark{3},
Manoj Joshi\altaffilmark{4,5}, 
Xi Zhang\altaffilmark{6}, and
Ravi Kumar Kopparapu\altaffilmark{1,2,7,8}}


\altaffiltext{1}{Blue Marble Space Institute of Science, 1001 4th Ave Suite 3201, Seattle, WA 98154, USA}
\altaffiltext{2}{NASA Astrobiology Institute's Virtual Planetary Laboratory, P.O. Box 351580, Seattle, WA 98195, USA}
\altaffiltext{3}{Laboratory for Atmospheric and Space Physics, Department of Atmospheric and Oceanic Sciences, University of Colorado, Boulder, Colorado, USA}
\altaffiltext{4}{School of Environmental Sciences, University of East Anglia, Norwich Research Park, Norwich NR4 7TJ, UK}
\altaffiltext{5}{Climatic Research Unit, University of East Anglia, Norwich Research Park, Norwich NR4 7TJ, UK}
\altaffiltext{6}{Earth and Planetary Sciences Department, University of California, Santa Cruz, California, USA}
\altaffiltext{7}{NASA Goddard Space Flight Center, 8800 Greenbelt Road, Mail Stop 699.0 Building 34, Greenbelt, MD 20771, USA}
\altaffiltext{8}{Department of Astronomy, University of Maryland, College Park, MD 20742, USA}


\begin{abstract}
We investigate the atmospheric dynamics of terrestrial planets in synchronous rotation within the habitable zone of low-mass stars using the Community Atmosphere Model (CAM). The surface temperature contrast between day and night hemispheres decreases with an increase in incident stellar flux, which is opposite the trend seen on gas giants. We define three dynamical regimes in terms of the equatorial Rossby deformation radius and the Rhines length. The slow rotation regime has a mean zonal circulation that spans from day to night side, with both the Rossby deformation radius and the Rhines length exceeding planetary radius, which occurs for planets around stars with effective temperatures of 3300 K to 4500 K (rotation period $>20$ days). Rapid rotators have a mean zonal circulation that partially spans a hemisphere and with banded cloud formation beneath the substellar point, with the Rossby deformation radius is less than planetary radius, which occurs
for planets orbiting stars with effective temperatures of less than 3000 K (rotation period $<5$ days). In between is the Rhines rotation regime, which retains a thermally-direct circulation from day to night side but also features midlatitude turbulence-driven zonal jets. Rhines rotators occur for planets around stars in the range of 3000 K to 3300 K (rotation period $\sim5$ to $20$ days), where the Rhines length is greater than planetary radius but the Rossby deformation radius is less than planetary radius. The dynamical state can be observationally inferred from comparing the morphology of the thermal emission phase curves of synchronously rotating planets. 
\end{abstract}


\keywords{planets and satellites: atmospheres, planets and satellites: terrestrial planets, stars: low-mass, astrobiology}

\section{Introduction}

M-dwarf stars provide an abundance of environments for potentially hosting habitable planets. 
The discoveries of Proxima Centauri b around our closest stellar neighbor \citep{anglada2017} and 
the seven planets of the TRAPPIST-1 system \citep{gillon2017} indicate that M-dwarfs can harbor terrestrial 
plants within their liquid water habitable zones \citep{kasting1993,selsis2007,kopparapu2013,kopparapu2014,yang2013,yang2014}, which makes them likely candidates
for upcoming surveys with \textit{JWST} and \textit{TESS}.
Due to the small size of their host stars, and their short period orbits, habitable planets around M-dwarf stars are optimal targets
for detection and characterization of their atmospheres.

Speculation that planets in orbit around low-mass stars would be prone to synchronous rotation---so that one 
side experiences permanent day, while the other experiences permanent night---initially raised concern 
that such planets would be prone to freeze out their atmospheres and thus might not be habitable at all \citep{dole1964}.
But subsequent investigations with simplified climate models \citep{haberle1996} and general circulation models (GCMs) 
\citep{joshi1997,joshi2003,merlis2010,edson2011,showman2010,showman2013,leconte2013,yang2013,yang2014,cullum2014,hu2014,carone2014,carone2015,carone2016,wordsworth2015,way2016,kopparapu2016,kopparapu2017,noda2017,fujii2017}
have demonstrated that energy transport from the day to night hemisphere is generally sufficient to avoid atmospheric collapse
across a wide range of atmospheric compositions and rotation periods. 

Further analysis has revealed common patterns in the large-scale dynamics of synchronously rotating terrestrial planets, 
most notably a transition between circulation regimes as a planet's rotation period decreases and the 
Rossby deformation radius approaches planetary radius 
\citep{merlis2010,edson2011,showman2010,showman2013,leconte2013,yang2013,yang2014,haqqmisra2015,carone2014,carone2015,carone2016,noda2017}.
For an Earth-size synchronously rotating planet, this transition occurs at a rotation period of $\sim$5 days \citep{edson2011,carone2015}.
Recently, \citet{noda2017} explored the dependence of large-scale dynamics on a wide range of rotation periods and identified four
distinct dynamical regimes for synchronously rotating terrestrial planets, some of which are not explained by changes in the Rossby
deformation radius alone.

In this paper we discuss the dynamical regimes that characterize the atmospheres of habitable moist terrestrial planets in synchronous rotation around 
M-dwarf stars. Using the simulations conducted by \citet{kopparapu2017}, we examine the temperature contrast between the day and night side 
as these planets move toward the inner edge of the habitable zone. We then define three distinct dynamical regimes based upon
the equatorial Rossby deformation radius and the Rhines length, which define the most salient features of a planet's large-scale atmospheric 
dynamics. Such dynamical states could potentially be distinguished in future missions through observations of thermal emission phase curves.

\section{Model Description}

The set of GCM simulations by \citet{kopparapu2017} represent Earth-sized terrestrial 
planets with 1-bar N$_2$ atmospheres, where water vapor is the only greenhouse gas. 
Planets are assumed to be aquaplanets covered in a swamp ocean; thus, water is abundant in the system, limited only by the Clausius-Clapeyron relation.
Calculations are performed at increasing stellar flux up to the inner edge of the habitable zone, where the model atmosphere becomes
unstable with the initiation of a runaway greenhouse. Simulated planets have global mean surface temperatures ranging between $\sim$250-310 K.
This set of calculations is conducted with six different 
spectral energy distributions representing a range of M-dwarf host stars, with effective temperatures 
of 4500 K, 4000 K, 3700 K, 3300 K, 3000 K, and 2600 K using the BT-SETTL
grid of models \citep{allard2007}.

We assume all planets are in synchronous rotation with their host stars. This implies that the rotation period and orbital period
must be equal, which we calculate self-consistently for each case using Kepler's third law \citep{wordsworth2015,kopparapu2016,kopparapu2017} as
\begin{equation}
P=\left[\left(\frac{L_{\star}}{L_{\sun}}\right)\left(\frac{F_{\Earth}}{F_{\star}}\right)\right]^{\frac{3}{4}}\left[\frac{M_{\sun}}{M_{\star}}\right]^{\frac{1}{2}}\label{eq:kepler}.
\end{equation}
Here $P$ is the orbital (and rotational) period of the planet in years, $L_{\star}/L_{\sun}$ is the luminosity of the host star scaled by the luminosity of the sun, 
$F_{\star}/F_{\Earth}$ is the incident stellar flux on the planet scaled by the incident solar flux on Earth, and $M_{\star}/M_{\sun}$ is the mass of the host star 
in solar mass units. 

These simulations were all conducted with a modified version of the Community Atmosphere Model (CAM) from the National Center for 
Atmospheric Research (NCAR) in Boulder, Colorado. This version of CAM includes updates to the native radiative transfer by 
implementing a new correlated-\textit{k} method based on the HITRAN 2012 database, as well as increases to the infrared spectral resolution.
The radiative transfer in this version of CAM is valid for N$_2$-H$_2$O atmospheres with surface pressures up to 10 bar, and the 
GCM dynamical core includes the contribution of condensing water vapor to the surface pressure tendency. For a detailed discussion 
of this implementation of CAM, see \citet{kopparapu2017}.

\section{Day-Night Surface Temperature Contrast}\label{sec:daynight}

\begin{figure}
\epsscale{1.0}
\plotone{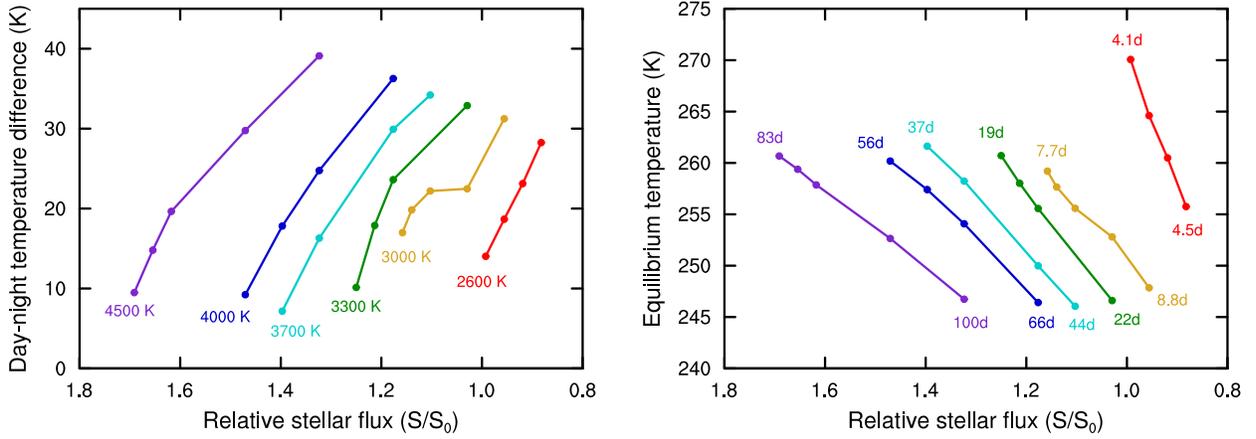}
\caption{The full set of simulations is shown as a function of relative stellar flux, $S/S_{0}$, 
with labels indicating the stellar effective temperature of the host star (left) and the rotation period of the planet (right),
and warmer stars at the left side of the panel.
As the model atmosphere warms due to an increase in $S/S_{0}$, the day-night temperature difference, $T_{day}-T_{night}$, decreases (left), while 
equilibrium temperature, $T_{eq}$, increases (right). 
\label{fig:eqT}}
\end{figure}

In order to detail 
the effectiveness of day-night energy transport across our simulation set, we examine the surface temperature difference between the day side
and night side hemispheres. Let $T_{day}$ be the area-weighted surface temperature
over the day side (substellar) hemisphere and $T_{night}$ be
the area-weighted surface temperature over the night side (antistellar)
hemisphere. 
We plot the day-night temperature difference in the left panel of Fig. \ref{fig:eqT} versus
relative stellar flux, which shows
that the day-night temperature difference decreases as a planet warms.
Fig. \ref{fig:eqT} includes markers at each of the GCM simulations conducted, labels for each stellar spectral type 
(left panel), labels for rotation period (right panel), and stellar flux scaled relative to the present-day solar flux, $S_0$. 
These simulations represent climate states ranging from the middle to inner edge of the habitable zone
for 1-bar N$_2$-H$_2$O atmospheres in synchronous rotation.

Previous studies using gray (\textit{i.e.}, non-wavelength-dependent) radiative transfer have suggested that the day-night temperature difference
shows no dependence upon rotation period for cloud-free atmospheres \citep{merlis2010,noda2017}.
Other studies have demonstrated that the presence and pattern of clouds depends on rotation rate, which 
alters the pattern of incident stellar radiation absorbed at the surface \citep{yang2014,kopparapu2016,kopparapu2017}.
Such experiments that separate the effects of incident stellar flux, stellar spectrum, and rotation rate are instructive 
for improving theoretical understanding of planetary atmospheres; however, 
Eq. (\ref{eq:kepler}) states that stellar flux and rotation period are inseparably linked 
for any synchronously rotating planets that are observed. Our goal with this paper is an attempt
to apply the knowledge gained from these previous theoretical investigations toward
the astronomically self-consistent set of GCM simulations by \citet{kopparapu2017}.

The left panel of Fig. \ref{fig:eqT} shows that at a fixed value of relative stellar flux, 
planets around hotter stars have a larger value of day-night temperature difference. This increased day-night temperature difference 
occurs due to changes
in both rotation period and the spectral energy distribution. These two effects are difficult to separate and can only be accounted for in a GCM 
with non-gray (\textit{i.e.} wavelength-dependent) radiative transfer. Although rotation period by itself is insufficient to explain the 
changes in day-night temperature difference, it remains a contributing factor when comparing synchronously rotating terrestrial planets 
around stars with different effective temperatures.

The right panel of Fig. \ref{fig:eqT} shows that equilibrium temperature, $T_{eq}$, increases as 
relative stellar flux increases,
where $T_{eq}^{4}=S(1-\alpha)/4\sigma$ (where $S$ is stellar flux, $\alpha$ is top-of-atmosphere albedo,
and $\sigma$ is the Stefan-Boltzman constant). 
Equilibrium temperature is 
a measure of the energy balance of
planet, 
with the contributions of dynamical and physical processes to radiation balance
(such as clouds and surface albedo) encapsulated in the planetary albedo parameter, $\alpha$.
We note that hotter stars have a lower equilibrium temperature than lower stars 
(when $S/S_0$ is constant), which occurs because cooler stars emit a lower percentage of visible 
radiation that can be absorbed at the planet's surface.
In our discussion that follows, we scale our temperature differences by $T_{eq}$ 
in order to show the relative contribution of direct stellar warming compared to atmospheric warming. 

\begin{figure}
\epsscale{1.0}
\plotone{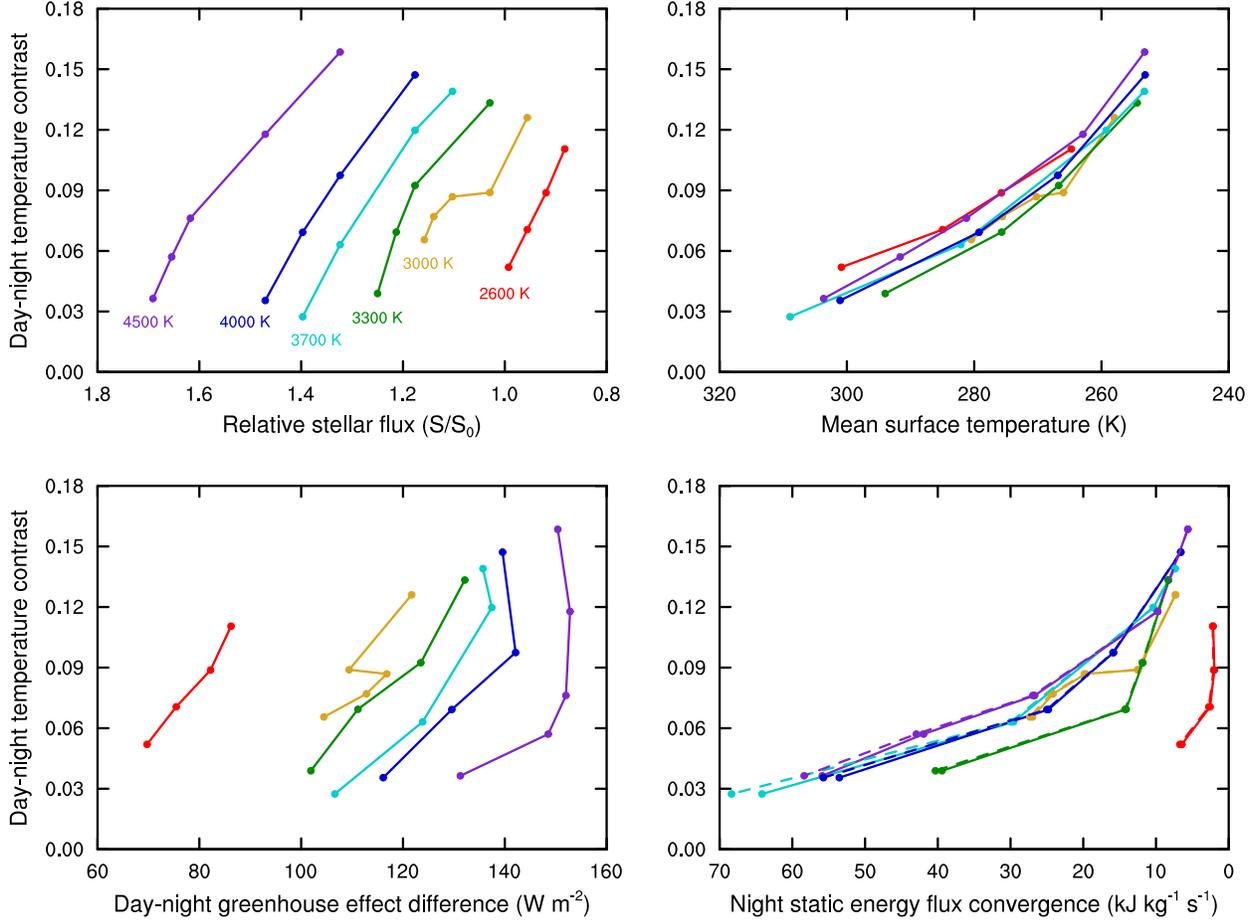}
\caption{The full set of simulations is shown as scaled day-night temperature contrast, $(T_{day}-T_{night})/T_{eq}$, versus
relative stellar flux, $S/S_{0}$, with warmer stars at the left side of the panel (top left). These simulations all 
show a correlation between scaled day-night temperature contrast and mean surface temperature (top right). 
This decrease in day-night temperature contrast as the planet warms corresponds to 
a decrease in the day-night greenhouse effect difference (bottom left), which occurs as a result of an increase in the 
the dry static energy flux convergence on the night side (bottom right, solid lines). 
The total static energy, with the moist latent energy component included, is also shown (bottom right, dashed lines), 
but the contribution of this latent energy is very small.
\label{fig:deltaT}}
\end{figure}

We define the day-night temperature contrast as $(T_{day}-T_{night})/T_{eq}$ (following the same approach as \citet{koll2016}), 
which provides a non-dimensional parameter for comparing simulations with different host stars.
We plot this day-night temperature contrast in the top left panel of Fig. \ref{fig:deltaT} as 
a function of relative stellar flux, where higher values represent greater day-night contrast.
The day-night temperature contrast (top left panel of Fig. \ref{fig:deltaT}) decreases as relative stellar flux increases, 
which shows similar functional morphology as the day-night temperature difference (left panel of Fig. \ref{fig:eqT}).

We next consider how the mean surface temperature in our simulations compares with the day-night temperature contrast. 
The top right panel of Fig. \ref{fig:deltaT} shows that all planets in our simulation set
exhibit a decrease in the day-night temperature contrast as the global mean surface temperature increases. All cases in the simulation 
set show a strong correlation between day-night temperature contrast and mean surface temperature, with the warmest simulations
showing the smallest day-night temperature contrast. (Here and elsewhere, we reverse the horizontal axis so that the warmest simulations are consistently at the left side of the panel.)
The decrease in day-night temperature contrast correlates with a 
rise in total greenhouse effect as the stellar flux increases, where total greenhouse effect, $F_{GH}$, is calculated 
as the difference between the surface outgoing longwave flux and top-of-atmosphere infrared flux, $F_{OLR}$, so that 
\begin{equation}
F_{GH} = {\sigma}T_s^4-F_{OLR}\label{eq:greenhouse}.
\end{equation} 
Eq. (\ref{eq:greenhouse}) suggests two possibilities for causing a change in the greenhouse effect as stellar flux increases.
Dry energy transport that yields net warming provides one mechanism that can increase $T_s$ and thus
increase $F_{GH}$. Moist processes provide a second mechanism, with the accumulation of water vapor (the only greenhouse 
gas in our simulations) causing a decrease in $F_{OLR}$ that likewise increases $F_{GH}$.
The bottom left panel of Fig. \ref{fig:deltaT}
shows that the difference in greenhouse effect between the day side and the night side decreases
as the planet warms. That is, as stellar flux increases for these planets,
the magnitude of greenhouse effect approaches equality between day and night hemispheres.

The dry contribution to the decrease in day-night temperature contrast 
can be explained by an increase in the 
static energy flux convergence on the night side. Static energy, $s$, is the sum of an air parcel's 
internal energy, $c_{p}T$, (\textit{i.e.}, enthalpy), potential energy due to gravity, $\Phi$, and latent energy due to moisture, $L_{v}q$. 
(Here $c_{p}$ is the heat capacity of air at constant pressure, $\Phi$ is geopotential, $L_{v}$ is the
enthalpy of vaporization of water, and $q$ is specific humidity.) 
The static energy flux, $\mathbf{v}s$, represents the change in static energy due to wind (where $\mathbf{v}$ is the horizontal wind vector),
while the static energy flux convergence, $-\nabla\cdot\left(\mathbf{v}s\right)$, 
describes the inward flow of static energy. 
An increase in static energy flux convergence on the night side will lead to an increase in internal energy, 
which, by Eq. (\ref{eq:greenhouse}), causes $F_{GH}$ to increase. 
Ultimately, this increase in static energy flux convergence 
occurs as a response to the deepening pressure contrast between day and night hemispheres 
as a synchronously rotating planet warms, which increases divergence on the day 
side and convergence on the night side of the component of wind 
known as the \textit{isallobaric wind} (see Appendix \ref{appendix:A} for additional discussion).
As these planets warm due to an increase in stellar flux, this increase in 
static energy flux divergence on the day side and convergence on the night side 
causes the temperature difference between the two hemispheres to decrease. 

The bottom right panel of Fig. \ref{fig:deltaT} shows 
the vertically-integrated static energy flux convergence per unit mass on the 
night hemisphere. This quantity increases as the day-night temperature contrast decreases, which 
corresponds to a similar increase in vertically-integrated static energy flux divergence on the day 
hemisphere (not shown). Solid lines in this figure show the flux convergence with only the
dry components of static energy ($c_{p}T+\Phi$), while dashed lines indicate the static energy flux convergence with moisture included ($c_{p}T+\Phi+L_{v}q$).
The dry static energy dominates the flux convergence, with minimal effects of latent heating
evident only in the warmest simulations. For most simulations, the moist (dashed) static energy flux 
convergence cannot be distinguished from the dry (solid) static energy flux convergence, which 
reveals that the contribution of the static energy flux convergence to the day-night temperature contrast
is an inherently dry phenomenon that does not necessarily depend upon latent heating from moisture.

Many of our simulations also feature a night side temperature inversion above the surface (up to about 800 hPa).
Such inversion layers also appear in other GCM simulations of synchronously rotating atmospheres \citep{joshi1997,merlis2010,leconte2013} 
and can be replicated in simpler radiative-convective subsiding models \citep{koll2016}.  
Note, similar temperature inversions are observed in the cold, dry, and dark polar winters on Earth \citep{curry1996,liu2006}, which
provide an analog to the dark antistellar hemispheres of synchronously rotating worlds.
In such cases the atmosphere radiates from the layer above this inversion, which provides additional infrared flux into the surface and can 
even contribute to a negative net greenhouse effect, $F_{GH}<0$, for some of the coldest simulations.   
The inversions, both for antistellar hemispheres and for polar winters on Earth, are maintained by vigorous radiative cooling of 
land and ice surfaces along with the transport of overlying warmer air masses in the free atmosphere \citep{bintanja2011}.

Unlike Earth's poles, the antistellar hemispheres of synchronously rotating planets always remains dark, and thus the 
inversion can be a permanent feature.  However, the night side inversion decreases in strength as the planet warms, with the 
destruction of the inversion triggered by the increase in dry static energy flux convergence on the 
night side near the surface (Fig. \ref{fig:deltaT}).  Warming of the night side surface layers from dry static energy convergence then 
increases water vapor abundance, which increases the night side water vapor greenhouse effect and causes further warming.  
Here, we find that the night side inversions vanish for planets with mean surface temperatures of $\ge$300 K. 
Water vapor increases in both hemispheres as stellar flux increases and the surface warms.  
While the absolute magnitude of greenhouse effect increases all across the planet as the atmosphere warms, the difference in the greenhouse effect between 
day and night side lessens, which contributes toward reducing the day-night temperature contrast.

\subsection{Comparison with Gray Analytic Theory}

The contributions from water vapor accumulation and dry energy transport to total greenhouse warming in Eq. (\ref{eq:greenhouse})
can be represented as an equivalent optical gray depth, which summarizes this combined warming in a parameter that can 
be compared with gray analytic theory. \citet{koll2016}, following a similar approach as \citet{wordsworth2015}, analyze the atmospheres of dry synchronously rotating 
planets by drawing upon Carnot's theorem to describe the scaling of surface 
winds as a heat engine. Using a gray analytic two-column model, 
\citet{koll2016} develop expressions for $T_{day}$ and $T_{night}$, 
\begin{equation}
T_{day}\approx2^{1/4}T_{eq}\left[1-\frac{3\tau_{LW}}{4\left(1+4\frac{R}{c_{p}}\right)}\right]\label{eq:Kollday}
\end{equation}
and
\begin{equation}
T_{night}\approx2^{1/4}T_{eq}\frac{\tau_{LW}^{1/4}}{\left(1+4\frac{R}{c_{p}}\right)^{1/4}},\label{eq:Kollnight}
\end{equation}
where $\tau_{LW}$ is the globally-averaged longwave optical depth and $R$ is the dry gas constant.
Eq. (\ref{eq:Kollday}) shows that the day side temperature decreases with $\tau_{LW}$, while Eq. (\ref{eq:Kollnight}) reveals
that the night side temperature increases with $\tau_{LW}$. 
Combining Eqs. (\ref{eq:Kollday}) and (\ref{eq:Kollnight}) to obtain $(T_{day}-T_{night})/T_{eq}$ shows that the day-night temperature contrast
decreases as $\tau_{LW}$ increases.
This remains consistent with our model results that show an increase in greenhouse effect, analogous
to an increase in $\tau_{LW}$, associated with the decrease in day-night temperature contrast.

\begin{figure}
\epsscale{1.0}
\plotone{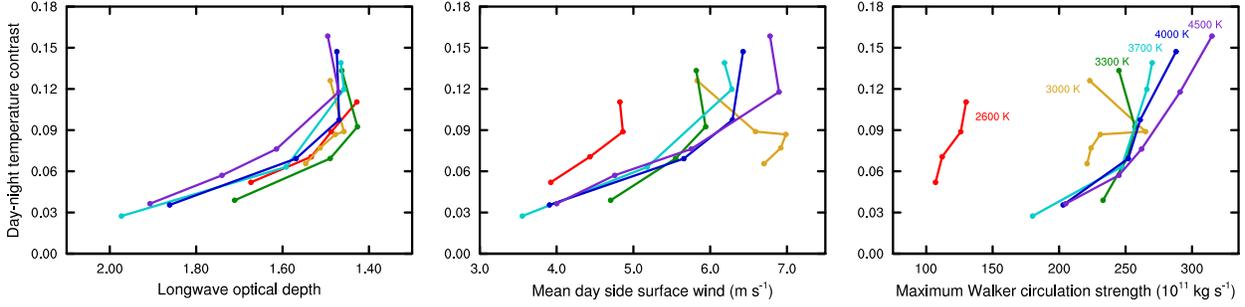}
\caption{The longwave equivalent gray optical depth calculated from Eqs. (\ref{eq:Kollday}) and (\ref{eq:Kollnight}) shows that these model atmospheres 
become increasingly optically thick as the planet warms and the day-night temperature contrast shrinks (left). 
The mean day side surface wind (middle) and maximum strength of the Walker circulation (right) both tend to 
decrease as the planet warms.
Note that the 2600 K and 3000 K cases show significant differences from the other simulations.
\label{fig:utau}}
\end{figure}

Substituting values of $(T_{day}-T_{night})/T_{eq}$ from our results into Eqs. (\ref{eq:Kollday}) and (\ref{eq:Kollnight}) allows us to calculate 
an equivalent gray optical depth for our set of GCM calculations, which we show in the left panel of Fig. \ref{fig:utau}. 
In our case, the parameter $\tau_{LW}$ represents any process that contributes to greenhouse warming (Eq. (\ref{eq:greenhouse})), 
which includes both the accumulation of water vapor and increase in static energy flux convergence as stellar flux increases, 
as well as cloud processes. 
Although the analytic model of \citet{koll2016} focuses on dry atmospheres, this relationship between $\tau_{LW}$ and $F_{GH}$
still remains qualitatively consistent with our non-gray, moist GCM results. 

The \citet{koll2016} heat engine analogy continues by predicting that mean day side 
surface wind, $U_s$, should scale as $U_s^3\sim\left(T_{day}-T_{eq}\right)\left(1-e^{-\tau_{LW}}\right)T_{eq}^4$.
This expression, and a similar one developed by \citet{wordsworth2015}, compares favorably with dry GCM
calculations that use gray radiative transfer; however, applying this relationship to our moist GCM results
by substituting from Eqs. (\ref{eq:Kollday}) and (\ref{eq:Kollnight}) suggests that $U_s$ should increase as the planet warms. 
However, Fig. \ref{fig:utau} (middle panel) shows $U_s$ calculated from our GCM simulations as the area-weighted root mean squared surface wind on the day hemisphere,
which shows that the simulations with the smallest day-night temperature contrast have the smallest value of mean day side wind. Fig. \ref{fig:deltaT} (upper left)
shows that these simulations with the smallest day-night temperature contrast also show the warmest mean surface temperatures. This implies
that $U_s$ in our simulations tends to decrease as a planet moves toward the inner edge of the habitable zone and day-night temperature contrast shrinks.
Likewise, the strength of the zonal overturning (\textit{i.e.,} Walker) circulation also tends to 
decrease as the planet moves toward the inner edge of the habitable zone (Fig. \ref{fig:utau}, right panel). 
These results from our moist GCM simulations are somewhat inconsistent with dry analytic theory, by showing
a decrease in wind speed and zonal circulation strength due to the changes in both stellar insolation
and planetary rotation period.

The theoretical expression of \citet{koll2016} remains valid when we examine our results
at a fixed value of day-night temperature contrast. Fig. \ref{fig:utau} shows that at a fixed value of 
day-night temperature contrast (\textit{e.g.,} $(T_{day}-T_{night})/T_{eq}=0.12$), the mean day side 
surface wind increases with stellar effective temperature (middle panel) as does the maximum Walker
circulation strength (right panel). 
Stars with a higher stellar effective temperature emit a higher
proportion of energy at shorter wavelengths, which corresponds to additional surface heating
on planets orbiting such stars. 

Water vapor absorption is one key feature present in our GCM that is absent in the gray analytic model of 
\citet{koll2016}. Stars with a lower stellar effective temperature have stronger emission at infrared wavelengths, 
which allows for greater absorption of this incoming radiation by water vapor in the atmosphere. This effect 
is evident in Fig. \ref{fig:utau}: 
planets orbiting the hottest stars are largely transparent to incoming stellar radiation, which causes greater 
surface warming and leads to the strongest circulation and surface winds. Likewise, planets orbiting cooler stars 
absorb a larger fraction of incoming radiation, which causes less direct surface warming and leads to a reduction in the maximum Walker circulation strength.

In general, the analytic expressions by \citet{koll2016} can adequately describe variations in $U_s$ for large
changes in $\tau_{LW}$ (such as comparing planets with fixed day-night temperature contrast around 
stars of different spectral type), but
this relationship breaks down when considering the smaller changes in surface pressure and wind
that occurs as a terrestrial planet is moved closer toward the inner edge of the habitable zone.

\subsection{Comparison with Gas Giants}

The day-night temperature contrast is one of the primary observable features of exoplanet atmospheres. 
The decreasing trend of day-night surface temperature contrast with increasing equilibrium temperature from our terrestrial 
simulations is opposite for synchronously rotating gas giant planets, where observations show that an 
increase in heating leads to an increase in the day-night temperature contrast \citep{perezbecker2013,komacek2016,komacek2017}.
The day-night temperature contrast observed on an optically thick gas giant atmosphere 
occurs at the emission level in the free atmosphere, whereas the flux emitted by an
optically thin terrestrial atmosphere primarily emerges from the surface. 
These opposite trends imply that the two types of atmospheres are located in two fundamentally different regimes: 
a hot, dry, top-heated gas giant regime versus a cold, moist, bottom-heated terrestrial regime. 

For both of these regimes, the day-night heat transport in the free atmosphere of synchronously rotating planets is strongly influenced 
by radiation and zonally propagating waves \citep{showman2013,perezbecker2013,wordsworth2015,koll2015,koll2016,komacek2016,komacek2017,zhang2017}.
The free-atmosphere day-night 
temperature contrast can be predicted based on a scaling theory developed by \citet{komacek2016,komacek2017} and \citet{zhang2017},
\begin{equation}
\frac{T_{day}-T_{night}}{T_{eq}}\Big{|}_{free}\sim1-\frac{2}{\alpha+\sqrt{\alpha^2+4\gamma^2}},\label{eq:tfree}
\end{equation}
where the non-dimensional parameters $\alpha$ and $\gamma$ are defined as:
\begin{align}
\alpha&=1+\frac{\left(\Omega+\frac{1}{\tau_{drag}}\right)\tau_{wave}^2}{\tau_{rad}\Delta\ln p},
\\
\gamma&=\frac{\tau_{wave}^2}{\tau_{rad}\tau_{adv,eq}\Delta\ln p}.
\end{align}
Here $\Omega$ is rotation rate, $\tau_{rad}$ is the radiative timescale, $\tau_{wave}$ is the 
timescale for wave propagation, $\tau_{drag}$ is the frictional drag timescale, 
$\Delta\ln p$ is the thickness of the photosphere in terms of log pressure, and $\tau_{adv,eq}$ 
is the advective timescale due to the ``equilibrium cyclostrophic wind''. (See Appendix A from \citet{zhang2017} for more discussion.)

The key parameters here are the radiative timescale, $\tau_{rad}$, and wave propagation timescale, $\tau_{wave}$. 
The typical radiative timescale on a canonical hot Jupiter is about $10^4-10^6$ s in the photosphere \citep{showman2002}. 
The gravity wave speed, $(gH)^{1/2}$ where $g$ is gravitational acceleration and $H$ is scale height, is on the order of 1 km s$^{-1}$, resulting a wave propagation 
timescale of $10^4-10^5$ s, which is comparable to the radiative timescale. However, the temperature dependence of the two timescales is different.
In the dry atmosphere of a gas giant, 
$\tau_{rad}\propto T_{eq}^{-3}$ \citep{showman2002} if opacity does not significantly change with temperature. On the other hand, in 
an isotherm limit, $\tau_{wave}\propto T_{eq}^{-1}$. As equilibrium temperature increases, the radiative timescale decreases more 
rapidly than the wave propagation timescale. Eq. (\ref{eq:tfree}) predicts that the day-night temperature contrast will increase. Physically, 
the atmosphere of a gas giant will become more radiation-controlled and the day-night heat transport will be less efficient by waves, which results in a 
larger day-night temperature contrast. Simulations by \citet{komacek2016,komacek2017} have confirmed this trend in the hot atmosphere regime for giant planets.

By contrast, due to their colder temperature and smaller planetary radius,
terrestrial atmospheres in the habitable zone generally show a much shorter wave propagation timescale
compared to the radiative timescale \citep{selsis2011,koll2015}.
For example, a typical radiative timescale on an Earth-like planet is a few days, but the gravity wave propagation 
timescale is about an hour due to its smaller planetary radius. Thus the ratio $\tau_{wave}/\tau_{rad}$ is on the order of 0.01 or less \citep{koll2015}.
This implies that the temperature of the day side and 
night side are homogenized in the free atmosphere, which leads to the ``weak temperature gradient" regime for terrestrial atmospheres 
\citep{wordsworth2015,koll2015,koll2016}.
Unlike the gas giant regime, which are primarily heated from the top, the atmospheres of optically thin
terrestrial planets are primarily heated from the bottom. 
Surface temperatures, both on the day and night side, are approximately 
in radiative-convective equilibrium with the overlying atmospheres \citep{koll2016}. As a result, the day-night surface temperature contrast 
in the terrestrial regime is primarily governed by the opacity that controls the surface-atmosphere radiative flux exchange, 
as shown by Eqs. (\ref{eq:Kollday}) and (\ref{eq:Kollnight}). Given a particular stellar flux, increasing the opacity will tend to decrease the day side 
surface temperature but increase the night side surface temperature \citep{koll2016}, which therefore decreases the day-night 
temperature contrast at surface. If opacity does not change significantly with equilibrium temperature in the dry atmosphere, 
the day-night temperature contrast will be approximately constant for all planets. 
In thin moist atmospheres, the presence of water vapor, and 
its associated increase in longwave opacity with equilibrium temperature, leads to a decreasing trend of day-night surface temperature contrast.

\section{Dynamical Regimes}

We analyze three regimes for large-scale atmospheric dynamics on synchronously rotating planets---\textit{slow rotators}, 
\textit{rapid rotators}, and \textit{Rhines rotators}---and their dependence on the Rossby deformation radius and the Rhines length.
We define and discuss the relevant dynamical parameters for each of these cases below, which we 
compare with previous GCM studies that have examined the dynamical regimes of synchronously rotating planets. 
We show that terrestrial planets in the habitable zone of stars with an effective stellar temperature of 3700 K to 4500 K are in the slow rotation regime, while those around 2600 K stars 
are in the rapid rotation regime. Intermediate stellar cases in the range of 3000 K to 3300 K show a unique transitional dynamical state that 
we describe as the Rhines rotation regime. 

\begin{figure}
\epsscale{0.7}
\plotone{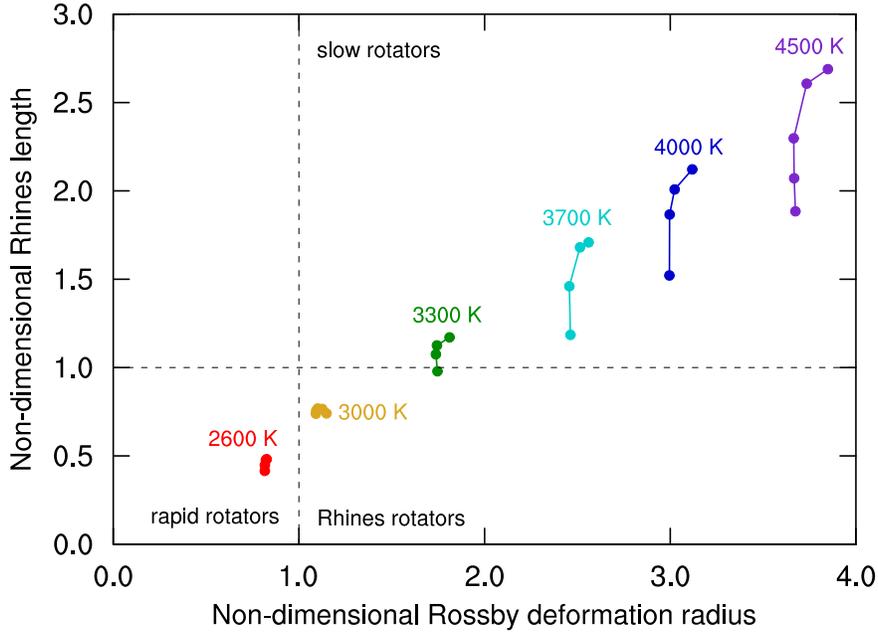}
\caption{Synchronously rotating planets are in the rapid rotation regime when the non-dimensional Rossby deformation radius is less than one, $\lambda_{R}/a < 1$,
which includes all 2600 K simulations. The Rhines rotation regime occurs when the non-dimensional Rhines length is less than one
but the non-dimensional Rossby deformation radius is greater than one, $L_R/a < 1$ and $\lambda_R/a > 1$, which includes all 3000 K simulations 
and one 3300 K case. 
Planets in the slow rotation regime have both $\lambda_{R}/a > 1$ and $L_R/a > 1$, which describes the remaining simulations for stars 3300 K to 4500 K.
\label{fig:Rhinesquad}}
\end{figure}

These three dynamical regimes are defined using two parameters: the Rossby deformation radius, which constrains the maximum extent 
of the zonal overturning circulation, and the Rhines length, which determines the maximum extent of zonally-elongated turbulent structures. 
We summarize this approach for characterizing the atmospheric dynamics of a synchronously rotating planet in Fig. \ref{fig:Rhinesquad}.
Slow rotators are found in the upper-right quadrant of Fig. \ref{fig:Rhinesquad}, 
where the non-dimensional Rossby deformation radius and non-dimensional Rhines length are both greater than one. Rapid rotators occur in the lower-left quadrant of Fig. \ref{fig:Rhinesquad},
where both the non-dimensional Rossby deformation radius and non-dimensional Rhines length are less than one. Rhines rotators describe a transitional dynamical state, which occurs in the 
lower-right quadrant of Fig. \ref{fig:Rhinesquad} with a non-dimensional Rossby deformation radius greater than one but a non-dimensional Rhines length less than one. 
We apply these regimes below in our discussion of terrestrial planets near the inner edge of the habitable zone, but this classification scheme can 
be generalized to planets with less Earth-like atmospheres, including synchronously rotating giant planets.

\subsection{Slow Rotators}

Slow rotators are characterized by strong convective motion beneath the substellar point, with energy transport by 
Rossby and Kelvin waves from the day to night side of the planet. Atmospheric dynamics on a slow rotator are
broadly characterized by a thermally-direct circulation with heating and rising air on the day side, cooling and descending air on the night side, and the 
strength of the circulation limited by frictional dissipation in the boundary layer \citep{koll2016}. Slow rotators are 
equivalent to the `Type-I' circulation regime described by \citet{noda2017}, where the thermally-direct day-night circulation 
is the primary characteristic of the planet's large-scale dynamics.

Previous studies have demonstrated that maintaining this
hemispheric large-scale circulation requires that the Rossby deformation radius is greater than 
the planetary radius \citep{merlis2010,edson2011,showman2010,showman2013,leconte2013,yang2014,haqqmisra2015,carone2014,carone2015,carone2016,noda2017}.
The Rossby deformation radius represents the ratio of buoyancy to rotational forces. For synchronously rotating planets,
the Rossby deformation radius is proportional to the maximum extent of the mean zonal circulation from day to night side.
Because we are primarily concerned with the equatorial propagation of Rossby, Kelvin, and other waves, we 
focus on the equatorial Rossby radius of deformation, $\lambda_{R}$, which we express, following \citet{gill1982}, as:
\begin{equation}
\lambda_{R}^{2}=\frac{\sqrt{gH}}{2\beta},\label{eq:rossby}     
\end{equation}
where $H$ is atmospheric scale height, $\beta=2\Omega/a$ represents the Coriolis parameter at the equator, $a$ is planetary radius, and $g = 9.81$ m s$^{-2}$. 
We express scale height as $H = \bar{T_s}R/m_{air}g$, where $\bar{T_s}$ is global mean surface temperature and $m_{air} = 0.028$ kg mol$^{-1}$ is 
the molar mass of air \citep{edson2011}.
We define slow rotators as planets where $\lambda_{R}/a > 1$. Using a present-day Earth GCM, \citet{edson2011} noted that this definition 
requires that slow rotators must have a rotational period of $\sim5$ days or greater, for an Earth-size planet 
in synchronous orbit around a G-dwarf star. Other studies using more idealized GCMs \citep{carone2014,carone2015,carone2016,noda2017} 
have found comparable values for this limit on slow rotators. All planets in our set of simulations have a rotation period greater than 5 days, except for planets orbiting a 2600 K host star. This 
means that, aside from the coolest host star, all these cases should expect to show $\lambda_{R}/a > 1$. 

The Walker circulation for the set of simulations is shown in Fig. \ref{fig:omegaMZC},
where contours show the strength of the mean zonal circulation, and shading shows rising motion. 
This particular set of simulations was chosen because they all have an approximately constant value of day-night
contrast, where $(T_{day}-T_{night})/T_{eq}\approx0.7$. Fig. \ref{fig:omegaMZC}
confirms that $\lambda_{R}/a > 1$ for all cases from 4500 K to 3000 K, where 
the planet's large scale dynamics are characterized by a thermally direct 
zonal circulation that spans the day-night hemisphere. All simulations with $\lambda_{R}/a > 1$
show the presence of this hemispheric zonal circulation, although 
only those cases that also show a non-dimensional Rhines number (defined in section \ref{sec:rhines}) greater than one
are classified as slow rotators.

\subsection{Rapid Rotators}

Rapid rotators are characterized by weak convective motion beneath the substellar point, 
with a Rossby deformation radius less than planetary radius ($\lambda_{R}/a < 1$) (Fig. \ref{fig:Rhinesquad}).
Rapid rotators tend to show 
banded cloud formation beneath the substellar point \citep{yang2014,kopparapu2016} and a mean 
zonal circulation that only partially reaches from day to night side \citep{haqqmisra2015,kopparapu2017}
Rapid rotators are comparable to the `Type-IV' circulation regime described by \citet{noda2017},
with some dynamical features that resemble present-day Earth.

\begin{figure}
\epsscale{1.0}
\plotone{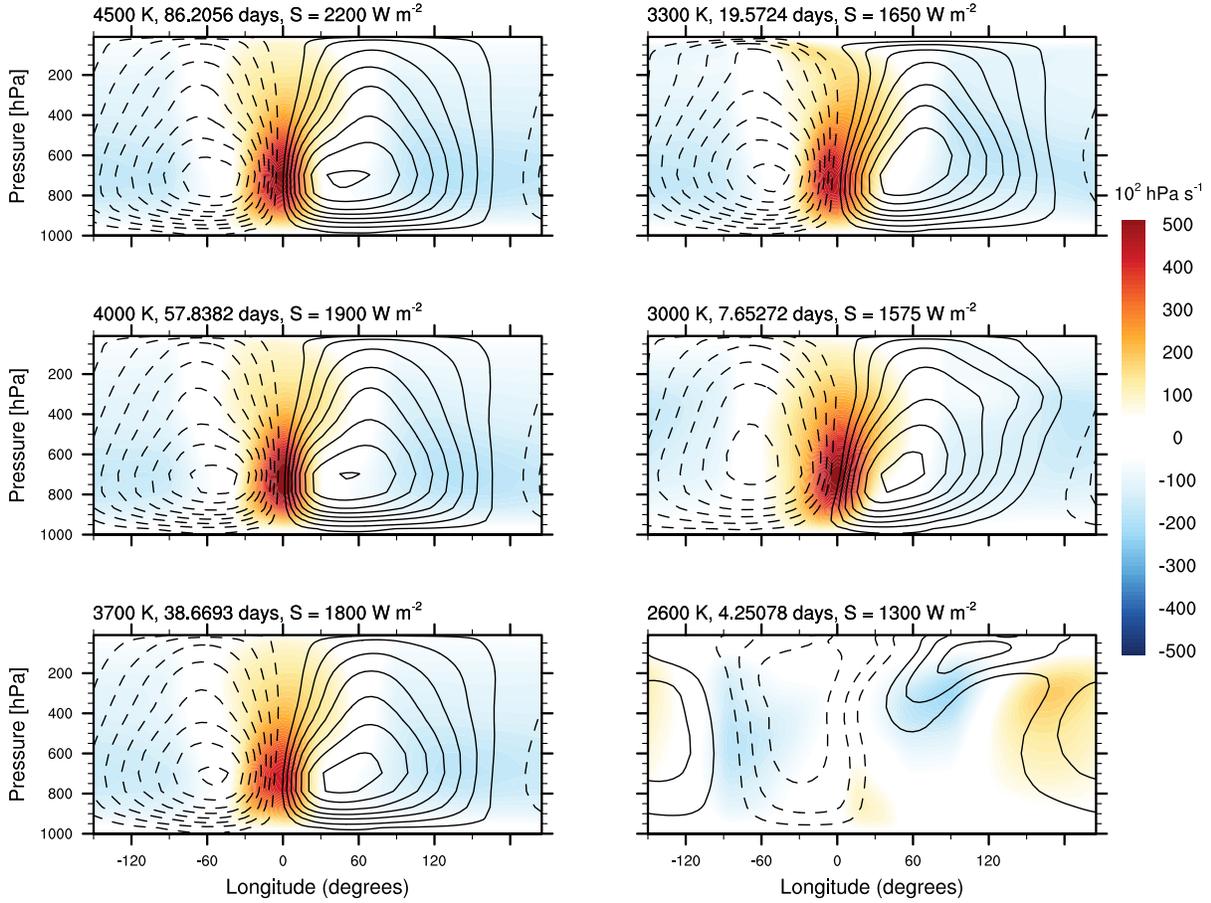}
\caption{The Walker circulation and meridionally-averaged vertical wind for synchronously rotating planets around 
4500 K, 4000 K, 3700 K, 3300 K, and 3000 K stars all show a thermally direct zonal circulation 
that spans day to night side. The 2600 K case shows rapid rotation with a Walker circulation that spans 
less than a full hemisphere. 
The contour interval for the Walker circulation is $30\times10^{11}\text{ kg s}^{-1}$,
with solid contours indicating clockwise circulation and dashed contours
indicating counterclockwise circulation. Shading indicates pressure
tendency, which corresponds to rising (warm colors) or sinking (cool
colors) motion.\label{fig:omegaMZC}}
\end{figure}

The 2600 K case appears as an outlier in the bottom right panel of
Fig. \ref{fig:deltaT}, which can be understood because these planets
are in a rapidly rotating regime with a zonal flow that
does not span the day-night hemisphere. 
This is shown in Fig. \ref{fig:omegaMZC},
where the 2600 K case exhibits its rapid rotation
through a less-organized zonal circulation with a larger number of
cells that do not span the day-night hemisphere. The 2600 K case also shows less
direct rising motion beneath the substellar point, which leads to 
a reduction in the amount of convectively-transported water vapor to the upper
troposphere.

\begin{figure}
\epsscale{1.0}
\plotone{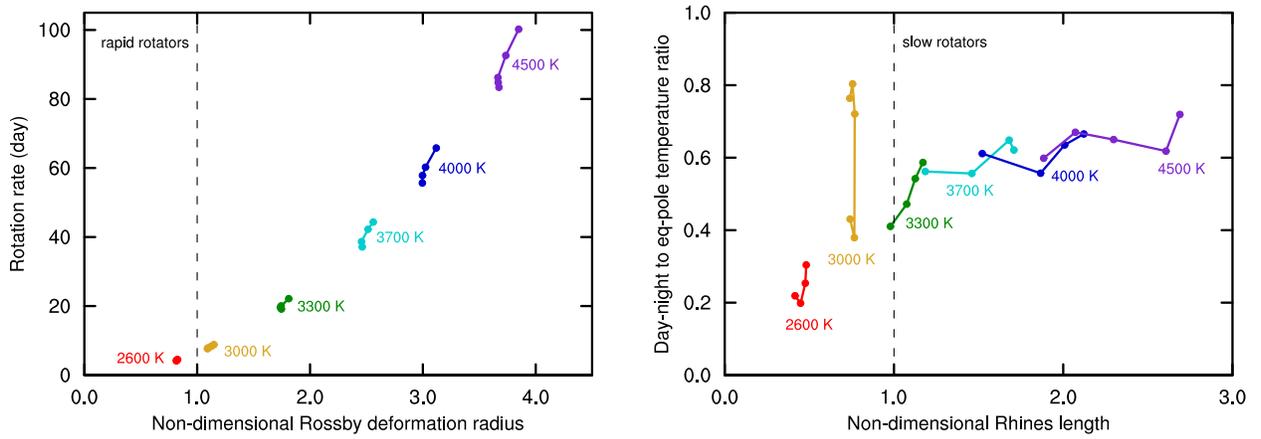}
\caption{Synchronously rotating planets are in the rapid rotation regime when the non-dimensional Rossby deformation radius is less than one, $\lambda_{R}/a < 1$,
which occurs at a rotation period of $\sim$5 days (left).
The Rhines rotation regime occurs when the non-dimensional Rhines length is less than one, $L_R/a < 1$ (right). Such planets respond 
to an increase in stellar flux by reducing the day-night temperature to equator-to-pole temperature ratio, $(T_{day}-T_{night})/(T_{equator}-T_{pole}$).
Planets in the slow rotation regime have both $\lambda_{R}/a > 1$ and $L_R/a > 1$.\label{fig:Rhines}}
\end{figure}

The set of GCM simulations is plotted in terms of rotation period versus $\lambda_R/a$ in Fig. \ref{fig:Rhines} (left panel), 
which shows that only the 2600 K case falls within the rapid rotation regime. The rest of the simulations
show a thermally-direct circulation that spans day to night hemisphere. Although the 3000 K case is close to the dashed 
line in the left panel of Fig. \ref{fig:Rhines} where $\lambda_R/a=1$, we can confidently exclude the 3000 K case
from being a rapid rotator, because its Walker circulation spans from day to night side (Fig. \ref{fig:omegaMZC}). Only when the rotation 
rate is less than $\sim$5 days (for Earth-mass planets) does the Walker circulation show evidence of the slow rotation dynamical 
regime.

Some studies have extended their GCM simulations to planets with a smaller rotation period of 1 day 
\citep{merlis2010,haqqmisra2015,carone2014,carone2015,carone2016,noda2017}. \citet{noda2017} even 
discuss a `Type-III' circulation regime that occurs at a rotation period of a few days, less than the rotational 
period of our 2600 K cases. We neglect these cases in our present study because Eq. (\ref{eq:kepler}) 
requires that any planets with a rotation period of 1 day or less that also resides in the habitable 
zone of its host star must therefore reside around very cool brown dwarf stars. Because our set of 
calculations extends only to 2600 K stars, such planets with short 1 day or less orbital periods are 
beyond the scope of this study.

\subsection{Rhines Rotators}\label{sec:rhines}

Rhines rotators are characterized by strong upper-atmosphere superrotation as well as strong
upwelling beneath the substellar point, occupying a transition region between slow and rapid rotators (Fig. \ref{fig:Rhinesquad}, lower-right quadrant).
Rhines rotators still show a Walker circulation that spans from day
to night side (Fig. \ref{fig:omegaMZC}), similar to slow rotators, but their atmospheric dynamics are also characterized by 
asymmetric zonal jets at midlatitudes. 
Rhines rotators are analogous to the `Type-II' circulation regime described by \citet{noda2017}, which
show the point of maximum heating beginning to drift off-center from the substellar point.
This transitional regime was also noted by \citet{edson2011} and \citet{carone2014}, which shows an 
increasing effect of equatorial superrotation dynamics. 

\begin{figure}
\epsscale{1.0}
\plotone{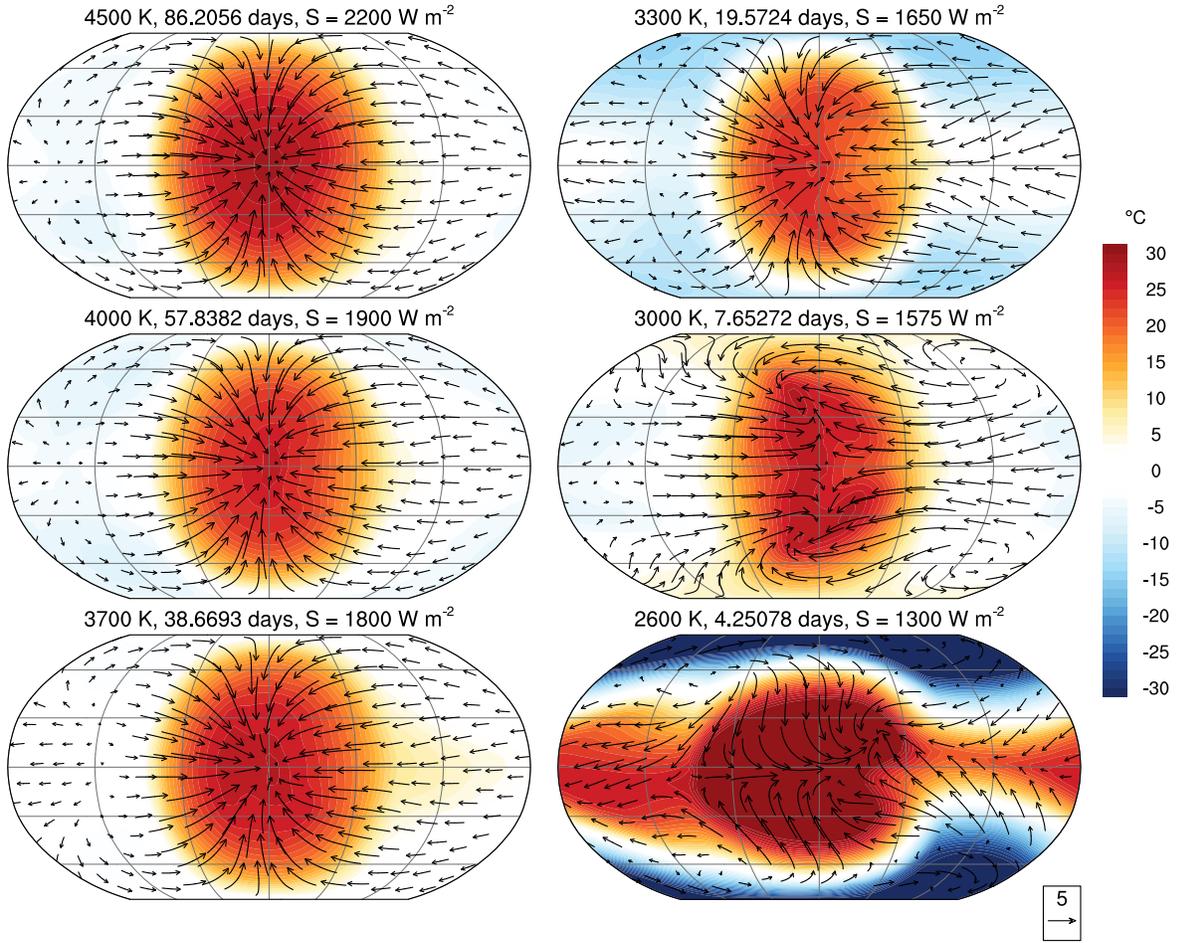}
\caption{Surface temperature and surface wind vectors show a nearly symmetric pattern of substellar heating and 
flow toward the substellar point for planets in the slow rotation regime around 4500 K, 4000 K, 3700 K, and 3300 K stars.
The 2600 K case in the rapid rotation regime shows asymmetric warming flow patterns that extend in an equatorial band 
from day to night side. In between is the Rhines regime, which shows departure from symmetry particularly at midlatitudes
for the 3000 K cases.\label{fig:surftemp}}
\end{figure}

Fig. \ref{fig:surftemp} shows surface temperature and wind vectors for the same set of GCM simulations 
as in Fig. \ref{fig:omegaMZC}. The 4500 K, 4000 K, and 3700 K cases all show characteristic flow inward 
toward the substellar point, with the surface flow in one hemisphere a mirror image of the other. This mirror symmetry
begins to break in the 3300 K case, with the beginnings of an eastward shift of the point of maximum heating. 
The 3000 K and 2600 K cases both show a notably different flow pattern along the surface and toward the substellar point, 
with turbulent features contributing to both the midlatitude and equatorial flow. In particular, the vortex-like structures
at midlatitude and polar latitudes of the 3000 K case in Fig. \ref{fig:surftemp} indicate the breaking of symmetry due
to turbulent flow. 

\begin{figure}
\epsscale{1.0}
\plotone{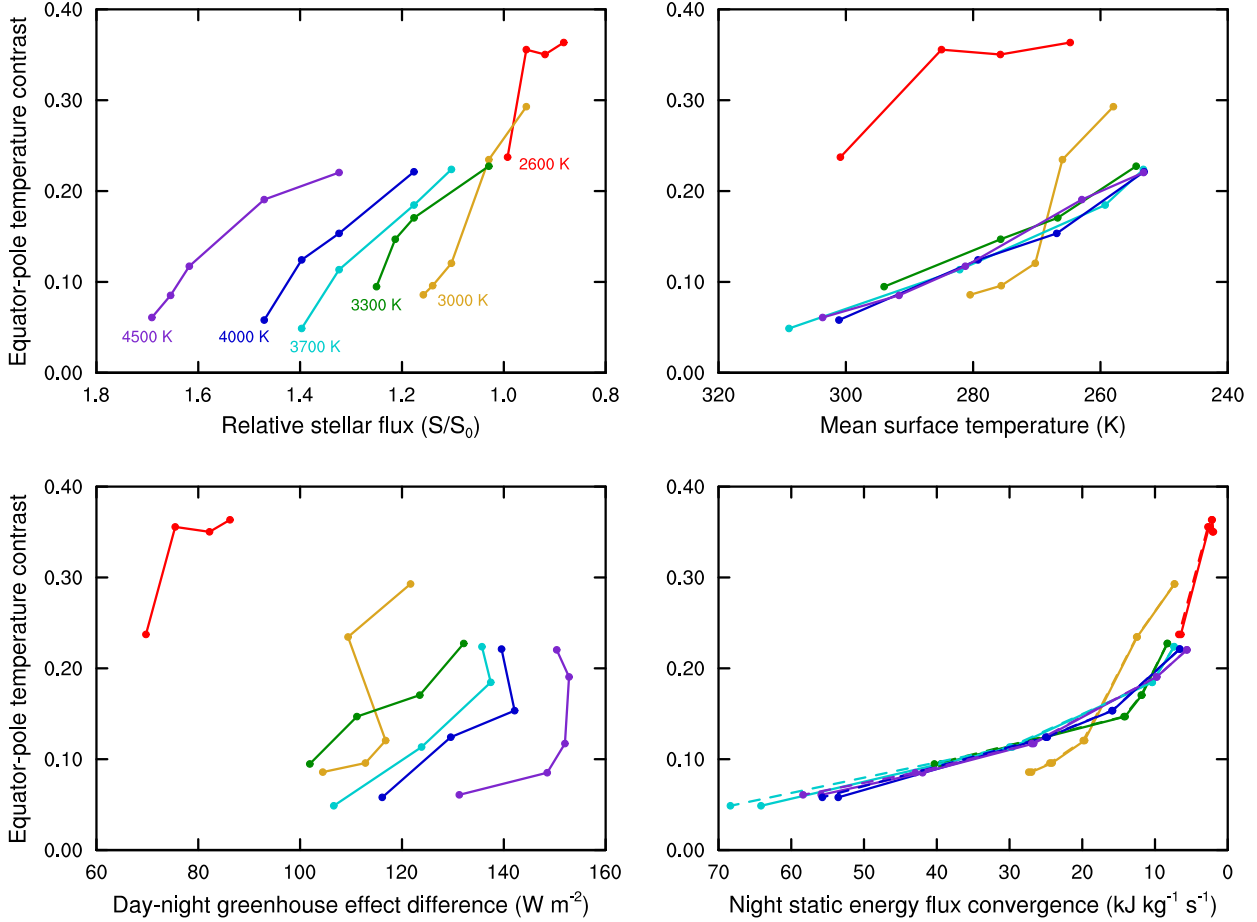}
\caption{The full set of simulations is shown as scaled equator-to-pole temperature contrast, $(T_{equator}-T_{pole})/T_{eq}$,
versus relative stellar flux, $S/S_{0}$, with warmer stars at the left side of the panel (top left). 
Simulations in the slow rotation regime show a correlation between scaled equator-to-pole temperature contrast and mean surface temperature (top right), 
which corresponds with the decrease in day-night greenhouse effect difference (bottom left) that occurs in response to the increase in night 
side dry (solid) and moist (dashed) static energy flux convergence (bottom right). 
The 2600 K rapid rotation and 3000 K Rhines rotation cases show notably different behavior due in part to their warmer poles compared to 
slow rotators.\label{fig:eqpole}}
\end{figure}

We further detail the energy distribution of our simulation set by examining the temperature 
difference between the substellar point at the equator and the pole. Let $T_{equator}$ be the maximum 
temperature beneath the substellar region and $T_{pole}$ be the minimum of the north and south pole temperatures.
Fig. \ref{fig:eqpole} shows the equator-to-pole temperature contrast
$(T_{equator}-T_{pole})/T_{eq}$ versus relative stellar flux, mean
surface temperature, the day-night greenhouse effect difference, 
and the vertically-integrated night side static energy flux convergence.
All the slow rotators from
4500 K through 3300 K show similar trends as Fig. \ref{fig:deltaT},
with the magnitude of equator-to-pole temperature contrast even greater
than the magnitude of day-night temperature contrast. By comparison, the 3000 K and
2600 K cases appear as outliers from the others in the two
panels on the bottom row of Fig. \ref{fig:eqpole}, 
which is due in part to their warmer poles compared to slow rotating cases.
This behavior
cannot be explained by the Rossby deformation radius alone, since
the 3000 K case shows $\lambda_{R}/a > 1$ while the 2600 K case is a rapid rotator.

Turbulent structures on Earth and other asynchronously rotating planets tend 
to elongate in the east-west direction compared to north-south, which \citet{rhines1975} 
realized is due to the variation in the Coriolis parameter with latitude. The latitudinal scale 
at which turbulent flow can organize into zonal jets is known as the Rhines length, which is defined as
\begin{equation}
L_{R}=\pi\sqrt{\frac{U}{\beta}}\label{eq:rhines},
\end{equation}
where $U$ is a characteristic root mean squared velocity at the relevant energy-containing 
scale \citep{showman2010,showman2013,vallis2017}. 
In general, the Rhines scale represents the transition scale between
turbulent and wave-driven motion. Turbulent structures on Earth can only grow to sizes limited by the 
Rhines scale, beyond which Rossby waves dynamics become the primary driver.
We define a non-dimensional Rhines length, $L_{R}/a$, which indicates the zonal 
scale to which turbulent structures can grow. 
On Earth and Jupiter today, $L_{R}/a<1$, thereby implying that any zonal jets that emerge
from turbulent energy cannot grow to encompass the entire planetary circumference. 
For synchronously rotating planets, we can use this non-dimensional Rhines length to determine 
the conditions under which turbulence-driven zonal jets can grow to planetary scales. For $L_{R}/a > 1$, 
the length scale for the turbulent energy cascade is greater than planetary radius, so we expect
atmospheric dynamics to be dominated by the thermally-direct circulation from day to night side, with
little contribution from turbulence-driven jets. Conversely, for $L_{R}/a < 1$, the length scale
for the turbulent energy cascade is smaller than planetary radius, so zonal turbulence-driven jets 
can form at midlatitudes and cause a departure from symmetry in surface flow.
We refer to these synchronously rotating planets, defined by $L_{R}/a < 1$ and $\lambda_{R}/a > 1$, as Rhines rotators.

Fig. \ref{fig:Rhines} (right panel) shows the ratio of the day-night temperature
contrast to the equator-to-pole temperature contrast ratio, $(T_{day}-T_{night})/(T_{equator}-T_{pole})$,
versus the non-dimensional Rhines number. 
Our calculation of the Rhines length assumes that $U$ is equal 
to the area-weighted root mean squared surface wind on the day hemisphere, as shown in Fig. \ref{fig:utau}.
For all the slow rotators from 4500 K to 3700 K, we see that $L_{R}/a > 1$, which implies
a large-scale dynamical structure characterized by a symmetric thermally-direct
circulation from day to night side (Fig. \ref{fig:omegaMZC}). 
By contrast, all cases for 2600 K and 3000 K stars show $L_{R}/a < 1$, which 
implies that the atmospheric dynamics of 
such planets include contributions at midlatitudes from turbulent-driven zonal jets
that break the symmetry of surface flow when compared to slow rotators.
Fig. \ref{fig:Rhines} also shows that the 3300 K simulation resides in
a transition zone between Rhines rotation and slow rotation, where $L_{R}/a \approx 1$.

Fig. \ref{fig:Rhinesquad} indicates the cases in our GCM simulations that fall within the Rhines rotation
regime, with $L_{R}/a < 1$ and $\lambda_{R}/a > 1$. These simulations all show very little spread in the value 
of $L_R$, even though stellar flux varies. 
The primary feature of planets within the Rhines rotation regime is that
the atmosphere responds to an increase in stellar energy by decreasing the day-night to equator-pole temperature contrast ratio (Fig. \ref{fig:Rhines}, right panel)
by breaking zonal symmetry without changing total kinetic energy by much. 
By comparison, planets in the slow rotation regime respond to an increase in stellar energy by decreasing 
the day-night temperature contrast as well as the root mean squared wind speed (Fig \ref{fig:utau}, right panel), which 
increases total kinetic energy.
Rhines rotators are able to develop zonal structures on the scale of the planetary radius, which induces the development 
of midlatitude and polar vorticies and other transient dynamical features that break the symmetry of the 
the thermally-directly model for large-scale circulation.

\begin{figure}
\epsscale{0.7}
\plotone{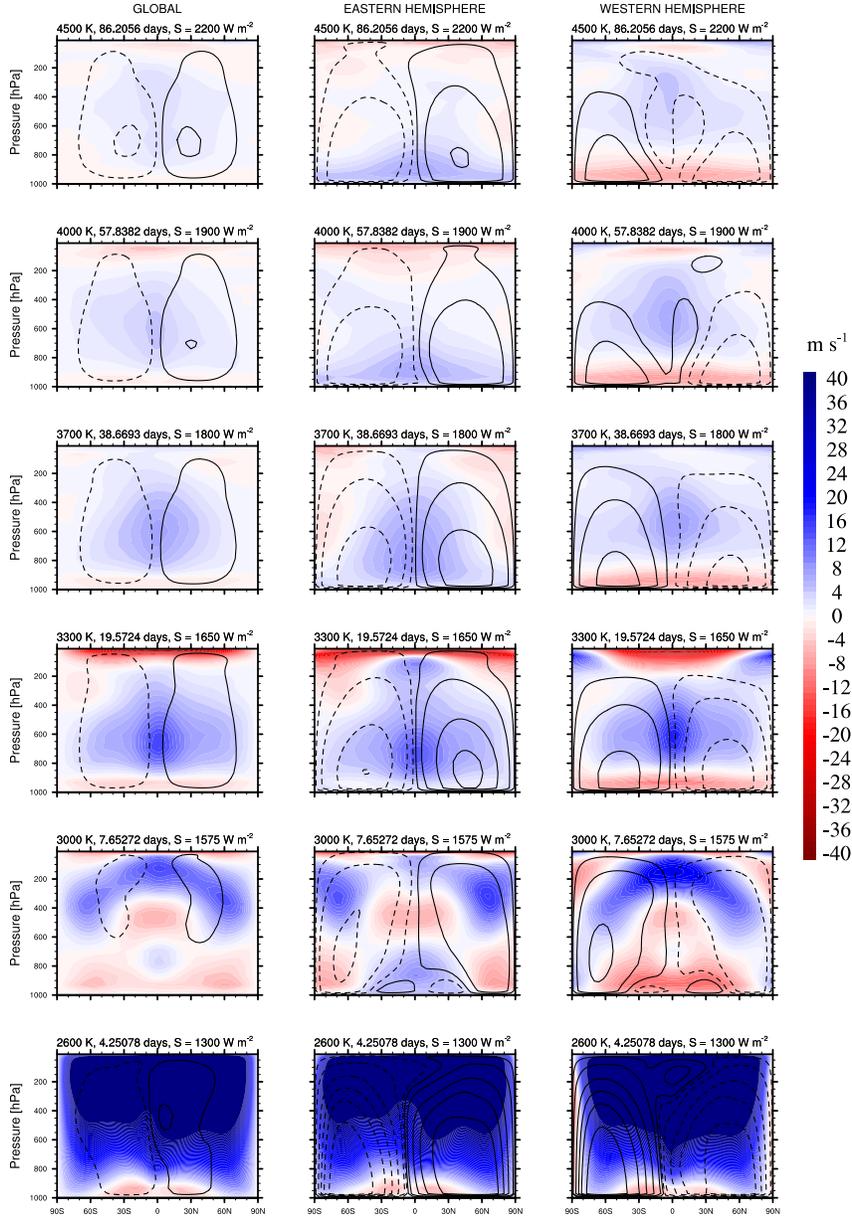}
\caption{The mean meridional circulation (line contours) and zonal mean zonal wind
(shading) are averaged across the entire planet (first column), eastern
hemisphere (second column), and western hemisphere (third column)
from the sub-stellar point. 
Note that all simulations show
circulation patterns in the opposite direction when comparing hemispheres.
Contours are drawn
at an interval of $\pm\{20,100,300,500,700,1000\}\times10^{9}\text{ kg s}^{-1}$.
Solid contours indicate clockwise circulation, and dashed contours
indicate counter-clockwise circulation. \label{fig:MMC}}
\end{figure}

We show zonal wind and the mean meridional (\textit{i.e.}, Hadley) circulation (MMC)
in Fig. \ref{fig:MMC} for the same set of cases as shown in Figs. \ref{fig:omegaMZC} and \ref{fig:surftemp}.
The left row shows the global quantities of zonal wind and MMC,
while the middle and right rows show the eastern and western hemisphere,
respectively, from the substellar point. Note that the direction of
the Hadley circulation changes sign when comparing the eastern and
western hemisphere, as discussed by \citet{haqqmisra2015}. This is a prediction
that emerges from the simplified shallow water study of \citet{geisler1981}, which 
found that the MMC changes direction on either side of a fixed heating source. 
This suggests that hemispheric separation of the MMC should always be examined 
for synchronously rotating planets, as the global mean MMC tends to cancel out 
branches with an opposite sign of circulation on either side of the substellar point.

The 4500 K through 3700 K cases all show surface flow into the substellar
point and a stronger easterly jet in the western hemisphere. By contrast
the 2600 K and 3000 K cases show strong upper-atmospheric superrotation
in both hemispheres. The 2600K slow rotators show very strong superrotation
at all latitudes and in both hemispheres, which owes to these planets
having both $L_{R}/a < 1$ and $\lambda_{R}/a < 1$.
The 3000 K case is a Rhines rotator with $L_{R}/a < 1$ and $\lambda_{R}/a > 1$, which 
manifests as zonal jets at the surface and middle of the atmosphere.
We see zonal propagation of Rossby waves dominating the
upper-atmosphere of the 3000K cases, although an upper-level westerly
jet also persists in these simulations. The 3300K case remains at
the threshold of the Rhines rotator transition, with a strong easterly jet persisting in both hemispheres and
a strong westerly jet at the top of the model atmosphere. 

The Rhines rotator regime represents a transitional, but distinct, dynamical state for 
terrestrial planets in synchronous rotation. Rhines rotators respond to an increase 
in stellar forcing by decreasing the equator-pole temperature contrast and are characterized 
by the presence of planetary-scale turbulent structures. Rhines rotators should be expected for 
terrestrial planets in the habitable zone of M-dwarf stars with an effective stellar temperature of 3000 K to 3300 K.

We also note that the upper-left quadrant of Fig. \ref{fig:Rhinesquad} represents 
a fourth regime where the Rhines length is larger than the Rossby deformation radius, although
this situation did not arise in our simulations. This regime could exist in principle, but most cold
atmospheres show a root mean squared velocity of the zonal wind less than the gravity wave speed (implying
that the Rhines length is smaller than the Rossby deformation radius). Hot atmospheres, such as closely-orbiting hot Jupiters, could 
conceivably host supersonic winds that may fall within this regime. However, this fourth regime is probably not feasible within
the terrestrial planet habitable zone.

\section{Implications for Observations}

All of our simulations show a strong correlation between the day-night surface temperature contrast and global mean surface temperature,
which is driven by the increase in dry static energy flux convergence on the night side (Fig. \ref{fig:deltaT}). 
This decrease in day-night contrast also can be interpreted as an increase in the atmospheric optical 
thickness, (Eqs. (\ref{eq:Kollday}) and (\ref{eq:Kollnight})), which suggests that synchronously rotating planets observed near the 
inner edge of the habitable zone should show a smaller day-night 
temperature contrast than planets orbiting farther outward. Planets near the outer edge of the habitable zone should build up a dense carbon dioxide
atmosphere \citep{kasting1993}, which also increases longwave optical depth and thus should decrease the day-night temperature contrast. 
Observations of the day-night temperature contrast on terrestrial synchronous rotators will provide important constraints on 
modeling the atmospheres of these planets.

The dynamical regime depends upon the Rossby deformation radius and Rhines length. Planetary rotation period 
can be inferred from observations of orbital period for planets around low mass stars where tidal locking is expected, 
but the combinations of models and observations will be needed to estimate values of mean wind speed, buoyancy, or other
indicators of the Rossby deformation radius and Rhines length. 
If we assume a fixed Earth-like radius, as with our set of simulations, then 
we can use the fact that rotation period is also a function of spectral stellar type (Eq. (\ref{eq:kepler})), which allows us to predict the dynamical 
state of planets in the habitable zone of a given low-mass star. Planets around M-dwarf stars with effective temperatures of 
3700 K to 4500 K (rotation period $>20$ days) should be slow rotators, with thermally-direct large-scale circulation from the day to night side. 
Other planets around the lowest mass stars we consider, with effective temperature of less than 3000 K (rotation period $<5$), should 
be rapid rotators that exhibit strong upper-level jets with asymmetric flow in the lower troposphere. Stars with effective 
temperatures in the range of 3000 K to 3300 K (rotation period $\sim5$ to $20$) represent the intermediate state of Rhines rotators, which exhibit planetary-scale turbulent flow 
while still retaining a thermally-direct circulation from the day to night side.

We can apply our model results to think about the expected dynamical regime for recently discovered terrestrial-sized exoplanets, assuming synchronous rotation. 
The TRAPPIST-1 systems may contain several planets within the traditional liquid water habitable zone, with TRAPPIST-1e being the most promising candidate \citep{wolf2017,turbet2017}.
With an orbital period of $\sim6$ days, TRAPPIST-1e would be close to the rapid rotator regime---especially if its lower planetary mass 
implies a lower atmospheric scale height (and thus a smaller value of $\lambda_R$). Proxima Centauri b has an orbital period of $\sim11$ days, which places 
it within the Rhines rotation regime if it is able to sustain an atmosphere. The planet LHS 1140b has a slower rotation period of $\sim24$ days, 
which implies that its surface flow and heating should exhibit the near-symmetry of the slow rotation regime. These conclusions 
not only assume that these planets are in synchronous rotation but also that the atmospheric dynamics can be approximated by our 
1-bar atmosphere simulations with an Earth-sized planet. 

These dynamical regimes can also be extended to planets with a larger radius than Earth. 
A larger planet has a faster rotation about its axis compared to a smaller planet with equal angular velocity.
For example, a planet with Earth-like terrestrial 
features but a radius twice as large as Earth would be a slow rotator with rotation period less than 10 days and a Rhines rotator 
with rotation period from $\sim$10 to 40 days. 
We also note that changes in atmospheric thickness and 
composition could also affect the Rossby deformation radius and Rhines number, and therefore the characterization of the atmosphere's dynamical state.
Further GCM studies that explore the mass and radius dependence of these dynamical regimes will place better constraints on known and future 
synchronously rotating exoplanets.

Thermal phase curves have been proposed as a relatively simple method for observing and characterizing terrestrial extrasolar 
planets (\textit{e.g.}, \citet{cowan2012}).  Thermal phase curves show the disk-integrated thermal flux emitted by a planet as 
seen by the observer as a function of the planet's position in its orbit around the host star.  Assuming an (approximately) 
edge on orientation, a phase angle of $\pm180^{\circ}$ corresponds with a transit event, when only the night side of the planet is in 
the field of view of the observer.  A phase of angle of $0^{\circ}$ corresponds with the secondary eclipse, where only the day side of 
the planet is visible to the observer (excluding the actual secondary eclipse event of course).  JWST will observe across most 
infrared wavelengths, and may be able to resolve the shape of thermal phase curves for terrestrial planets in nearby 
M-dwarf systems \citep{meadows2016}.

\begin{figure}
\epsscale{0.6}
\plotone{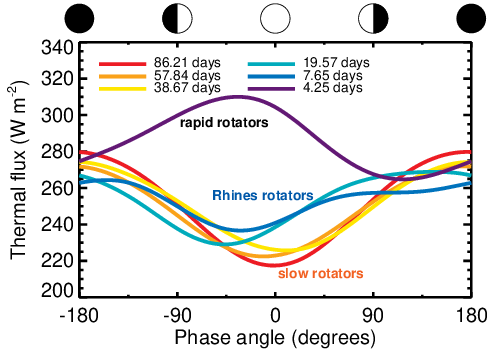}
\caption{Thermal phase curves can identify synchronously rotating habitable zone planets as slow rotators (minimum thermal flux near $0^\circ$), 
Rhines rotators (minimum thermal flux near $-45^\circ$), and rapid rotators (maximum thermal flux near $-45^\circ$).\label{fig:phasecurve}}
\end{figure}

We calculate broadband thermal emission phase curves following the method of \citet{koll2015}, using 
outgoing thermal flux maps produced by our GCM simulations, time averaged over many orbits.  In 
Fig. \ref{fig:phasecurve}, we show thermal phase curves for 6 different simulations.  Note that 
the simulations shown in Fig. \ref{fig:phasecurve} are the same set as shown in Figs. \ref{fig:omegaMZC}, \ref{fig:surftemp}, and \ref{fig:MMC}.  
Each of the three dynamical regimes are clearly evident and differentiable within broadband thermal emission phase curves. The differences 
between thermal phase curves for rapid, slow, and Rhines rotators are a result of the interaction of atmospheric dynamics with clouds and water vapor, 
which in turn feedback on both the planetary surface temperature and the allowed outgoing thermal radiation at the top of the atmosphere.  

\begin{figure}
\epsscale{1.0}
\plotone{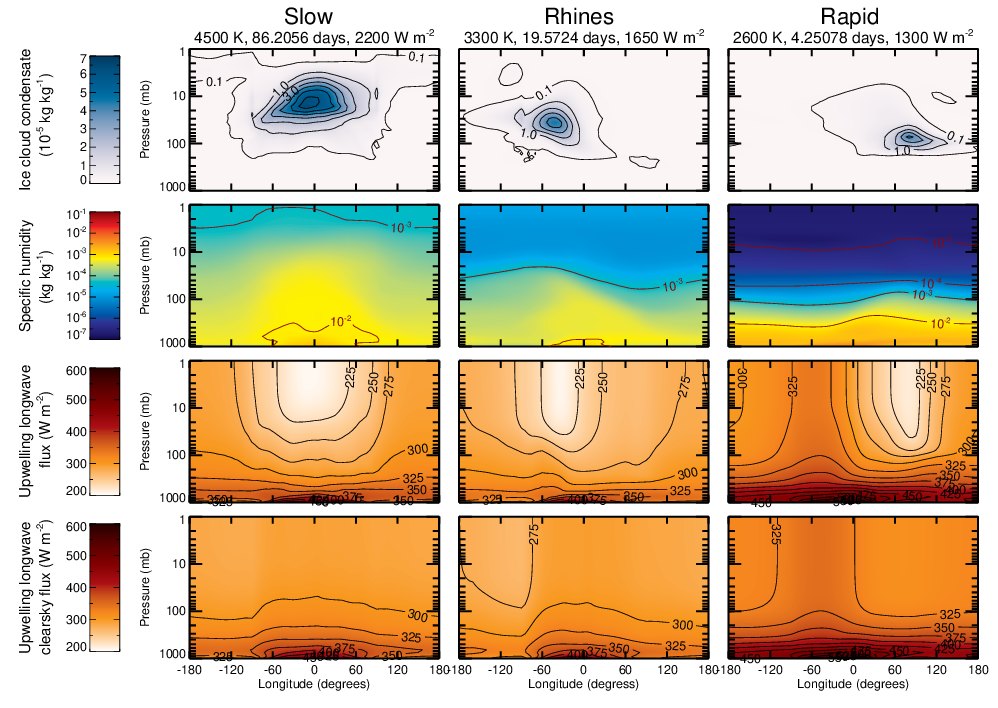}
\caption{Vertical profiles taken along the equator for ice cloud condensate mass mixing ratio (top row), specific humidity (second row), 
upwelling longwave flux (third row), and upwelling longwave clearsky flux (bottom row).  The substellar point is located at $0^{\circ}$ longitude.  
Ice clouds form in regions of high specific humidity and cold temperatures, typically above 100 mb.  The minimum in upwelling longwave flux corresponds 
to the maximum in ice cloud condensate for all simulations, because ice clouds are efficient greenhouse absorbers.  Slow rotators show areas of high 
specific humidity and cirrus clouds located immediately over the substellar point.  Water vapor, ice clouds, and the associated minima in the outgoing 
longwave flux are shifted westward of the substellar point for Rhines rotators and eastward of the substellar point for rapid rotators.  When clouds 
are ignored in the radiative transfer calculation, the upwelling longwave clearsky flux appears generally uniform across all longitudes.\label{fig:phaseclouds}}
\end{figure}

For each case, the modulation of the upwelling longwave flux (and thus the thermal emission phase curve) is tied closely to the location 
of high-altitude water ice (\textit{i.e.}, cirrus) clouds (Fig. \ref{fig:phaseclouds}).  Cirrus clouds are efficient absorbers of longwave radiation emitted 
by the planet surface \citep{ramanathan1989}.  Fig. \ref{fig:phaseclouds} shows the water ice cloud condensate mass mixing ratio, 
specific humidity, upwelling longwave flux, and upwelling longwave clearsky flux
for each dynamical regime, with all atmospheric columns located along the equator.  Note that the substellar point is at a 
longitude of $0^\circ$ and the antistellar point is a longitude of $\pm$180$^\circ$.
The bottom panel of Fig. \ref{fig:phaseclouds} shows that if we omit clouds from the radiative transfer calculation, then the upwelling longwave 
clearsky flux is generally uniform.  This indicates that cirrus clouds, rather than the advection of water vapor and its associated greenhouse effect,
are responsible for the morphology of the thermal phase curves.

Unsurprisingly, for all cases studied the hottest planetary surface temperatures are found on the day side (Fig. \ref{fig:surftemp}). 
In the absence of an atmosphere, or when the greenhouse effect is perfectly uniform everywhere, the thermal emission phase 
curve should mirror the disk integrated surface temperature distribution as a function of phase angle. 
However, cirrus clouds ultimately control where thermal energy can be effectively emitted to space.  Rapid rotators exhibit a maximum 
in thermal emission when viewing the day side of the planet.  Clouds on rapid rotating worlds are advected eastward off of the substellar 
point by the strongly super-rotating atmosphere.   A minimum in the thermal emission is coincident with the primary cirrus cloud deck 
that is located near the eastern terminator of the planet (Fig. \ref{fig:phaseclouds}).  Locations immediately west of the substellar point have relatively 
few clouds (\textit{e.g.}, \citet{kopparapu2017} Fig. 8), resulting in both warmer surface temperatures and efficient emission of thermal energy to space.  
Thus, the maximum in the thermal phase curve for rapid rotators is shifted $\sim$45$^\circ$ west of the substellar point (Fig. \ref{fig:phasecurve}), 
where surface temperatures are warm and clouds are absent.

For slow rotators again the surface temperature maximum is found at the substellar point; however thick and symmetric clouds completely enshroud 
the day side hemisphere.  A thick cap of high altitude cirrus clouds significantly reduces the outgoing thermal flux from the day side by lowering 
the emitting temperature of the atmosphere (Fig. \ref{fig:phaseclouds}).  The night side remains cloud free and can efficiently radiative energy to space.  
The night side thermal emission remains large despite cold surface temperatures, due to the near-surface inversion layer described in section \ref{sec:daynight}
and the low water vapor concentration, which allows the night side to act as a `radiator fin' that emits excess thermal energy to space
similar to the dry subtropics of Earth \citep{yang2014b}.
This results in a minimum in the thermal phase curve when viewing the day side of the planet and maximum when viewing the night side for slow 
rotators, as first noted by \citet{yang2013}.  Note that the thermal phase curve for slow rotators exhibits remarkable symmetry with minima 
at 0$^\circ$ and maxima at $\pm$180$^\circ$, due to the symmetric day-night general circulation that occurs on these worlds.
We also note that ocean heat transport will affect these results, as surface fluxes induces by the ocean can be significant
and can also depend upon planetary rotation period \citep{cullum2014,hu2014,way2016}, although 
the presence of continents could limit the magnitude of ocean heat transport \citep{yang2013}.

While differences between fast and slow rotators are evidently clear, the differentiation of Rhines rotators is more subtle. The thermal phase curve for Rhines 
rotators is qualitatively similar to that of slow rotators, however the minimum is shifted $\sim$45$^\circ$ westward of the substellar point.  Tropospheric water 
clouds still generally permeate the substellar hemisphere of Rhines rotators, keeping albedos large and suppressing the surface temperature.  However, 
the primary ice cloud layer is shifted westward of the substellar point, shifting the associated minimum in the thermal phase curve accordingly.  
This corresponds to the strong westerly jet present in the uppermost model layers of this Rhines rotator (Fig. \ref{fig:MMC}, middle row), which 
is nonexistent in the slow rotators and extremely weak in the rapid rotators. Rhines rotators may be differentiated 
from slow rotators by the offset of the minimum in the thermal phase curve.  Our results imply that understanding stratospheric 
process and ice cloud formation on terrestrial extrasolar planets may be critical for interpreting observed thermal phase curves, although
we caution that further studies with other GCMs, using different ice cloud parameterizations, will be needed to demonstrate robustness of these phase curve features.

\section{Conclusion}

Our examination of the atmospheres of planets in synchronous rotation around low mass stars reveals three distinct dynamical regimes (Fig. \ref{fig:Rhinesquad}). 
Rapid rotation occurs when the Rossby deformation radius is less than planetary radius (rotation period $>5$ days).
Slow rotation occurs when both the Rossby deformation radius and the Rhines length 
are greater than planetary radius (rotation period $>20$ days).
Rhines rotation occurs when the Rhines length is less than planetary radius but the Rossby deformation radius is greater than
planetary radius, which 
allows turbulent structures to reach planetary scales (rotation period $\sim5$ to $20$ days). 

These three dynamical states can be distinguished from one another through observations of the thermal phase
curve of the planet. Differences in the amplitude of the maxima and minima of the thermal phase curves
can be used to identify the slow rotation regime, while the transition between the rapid rotation and Rhines regimes 
can be identified by comparing the morphology of the thermal phase curves. Corroborating these phase curves with observations 
of the planet's orbital period and the host star's spectral type will provide a basis for further characterization 
of such atmospheres with computational models.

We also show that the day-night surface temperature contrast for terrestrial planets decreases as incident stellar flux increases.
The combined effects of moisture accumulation and the increase in static energy 
flux divergence on the night side leads to an increase in greenhouse effect that diminishes the temperature contrast between the day and night sides. 
We therefore expect that synchronously rotating planets near the inner edge of the habitable zone should
have diminished day-night surface temperature contrasts compared with other synchronous rotators at farther orbital distances.

We can apply these results to the response of a synchronously rotating atmosphere under the steady main sequence brightening of 
its host star.
Planets in the slow rotation regime respond to an increase in stellar forcing with a decrease in both the day-night temperature contrast and the root mean squared surface wind, 
whereas planets in the Rhines and slow rotation regimes respond to a similar increase by decreasing the day-night to equator-pole temperature contrast ratio.
This suggests that the atmospheres 
of slow rotators will adapt through reducing the equatorial day-night temperature contrast with an 
increase in static energy flux convergence on the night side. 
Rhines and slow rotators will react by increasing the equator-pole contrast more than the day-night contrast, tending toward a surface temperature distribution
with even heating at all latitudinal bands. It is beyond the scope of this paper to speculate as to which of these three regimes, 
if any, would be conducive or adverse to the presence of life. But we can at least begin to think about how the evolution of 
a low mass star can affect the dynamical state of a planet within its habitable zone.

\acknowledgments
The authors thank Jun Yang for constructive comments that greatly improved the manuscript.
J.H, E.T.W., and R.K.K. acknowledge funding from the NASA Habitable Worlds program 
under award NNX16AB61G. R.K.K. also acknowledges funding from NASA Astrobiology Institute's Virtual Planetary Laboratory lead 
team, supported by NASA under cooperative agreement NNH05ZDA001C. 
X.Z. acknowledges support from NSF grant AST1740921.
This work was facilitated though the use of advanced computational, storage, and networking infrastructure provided by the Hyak supercomputer system at the University of Washington.
This work benefited from the Exoplanet Summer Program in the Other Worlds Laboratory (OWL) at the University of California, Santa Cruz, a program funded by the Heising-Simons Foundation.
Any opinions, findings, and conclusions or recommendations expressed in this material are those 
of the authors and do not necessarily reflect the views of NASA or NSF.

\appendix
\section{Static energy flux convergence by isallobaric wind}\label{appendix:A}
As stellar flux increases, slow rotators and Rhines rotators
show a decrease in day-night temperature contrast that corresponds with an
increase in the night (day) side static energy flux convergence (divergence).
We demonstrate here that this relationship is ultimately driven by
increases in the night (day) side convergence (divergence) of the
component of wind driven by pressure contrasts known as the isallobaric wind.

Let $\mathbf{v}$ be the horizontal wind vector and $s=c_{p}T+\Phi+L_{v}q$
be the moist static energy per unit mass.
The static energy flux divergence can be written as 
\begin{equation}
\nabla\cdot\left(\mathbf{v}s\right)=s\nabla\cdot\mathbf{v}+\mathbf{v}\cdot\nabla s,\label{eq:energyflux}
\end{equation}
where the first term on the right is horizontal wind divergence, and
the second term on the right is advection of static energy. 
Static energy \textit{divergence} refers specifically to the situation when $\nabla\cdot\left(\mathbf{v}s\right) > 0$, 
while static energy \textit{convergence} implies that $\nabla\cdot\left(\mathbf{v}s\right) < 0$.
We are primarily concerned with comparing $\nabla\cdot\left(\mathbf{\mathbf{v}}s\right)$
between day and night sides, where the day side is characterized by
heating across much of the hemisphere and the night side is characterized
by near-uniform cold temperatures across most of the hemisphere. This
suggests that we can simplify Eq. (\ref{eq:energyflux}) by assuming
$\nabla s\approx0$ for these atmospheres.

The geostrophic wind describes the balance between the pressure gradient
force and Coriolis force, which we can write as 
\begin{equation}
\mathbf{v}_{g}=\mathbf{k}\times\frac{1}{\rho f}\nabla p,\label{eq:geostrophic}
\end{equation}
where $\mathbf{k}$ is the upward unit vector, $f$ is the Coriolis
parameter, $\rho$ is air density, and $p$ is pressure. However,
we cannot substitute $\mathbf{v}_{g}$ for $\mathbf{v}$ in Eq. (\ref{eq:energyflux})
because the geostrophic wind is non-divergent ($\nabla\cdot\mathbf{v}_{g}=0$).
We instead consider the quasigeostrophic approximation, where the
wind vector is the sum of geostrophic and ageostrophic components,
$\mathbf{v}=\mathbf{v}_{g}+\mathbf{v}_{a}$. The quasigeostrophic
approximation allows us to express the ageostrophic wind as
\begin{equation}
\begin{array}{ccc}
\mathbf{v}_{a} & = & \frac{1}{f}\mathbf{k}\times\frac{D\mathbf{v}_{g}}{Dt}\\
 & = & \frac{1}{f}\mathbf{k}\times\frac{\partial\mathbf{v}_{g}}{\partial t}+\frac{1}{f}\mathbf{k}\times\left(\mathbf{v}_{g}\cdot\nabla\mathbf{v}_{g}\right)
\end{array},\label{eq:ageostropic}
\end{equation}
following Holton (2004). The two terms on the bottom line of Eq. (\ref{eq:ageostropic})
respectively represent the isallobaric and advective components
of the ageostrophic wind. The advective term can contribute to ageostrophic
wind divergence through positive vorticity advection; however, vorticity
advection between day and night hemispheres is small. We therefore
focus our analysis on the contribution of the isallobaric wind to
the static energy flux. 

The isallobaric wind is perpendicular to lines of constant geopotential
tendency (known as isallobars) and describes flow toward regions of
falling pressure. Substituting from Eq. (\ref{eq:geostrophic}), we
write the isallobaric wind, $\mathbf{v}_{is}$, as

\begin{equation}
\mathbf{v}_{is}=\frac{-1}{f^{2}\rho}\nabla\left(\frac{\partial p}{\partial t}\right).\label{eq:isallobaric-1}
\end{equation}
We now replace the total wind vector in Eq. (\ref{eq:energyflux})
with the isallobaric wind in Eq. (\ref{eq:isallobaric-1}) to obtain
\begin{equation}
\nabla\cdot\left(\mathbf{v}s\right)\approx s\nabla\cdot\mathbf{v}_{is}=\frac{-s}{f^{2}\rho}\nabla^{2}\left(\frac{\partial p}{\partial t}\right).\label{eq:isallobaric}
\end{equation}
Eq. (\ref{eq:isallobaric}) shows that an increase (decrease) in pressure
corresponds to divergence (convergence) of the isallobaric wind. For
synchronous rotators, the increase in pressure due to fixed stellar
heating on a single hemisphere causes the isallobaric wind, as well
as the static energy flux, to diverge on the day side and converge
on the night side, which decreases the day-night temperature contrast.

\end{document}